\documentclass[%
 preprint,
 amsmath,amssymb,
 aps,
showkeys
]{revtex4-2}

\usepackage{graphicx}
\usepackage{dcolumn}
\usepackage{bm}
\usepackage{subcaption}
\usepackage{mathtools}
\usepackage{cases}

\begin{document}

\title{Tunneling half-lives in macroscopic-microscopic picture}

\author{Samyak Jain \\
ORCID: 0009-0000-7455-782X}

\affiliation{Department of Physics, Indian Institute of Technology Bombay, 
Powai, Mumbai 400 076, India}%
\email{Corresponding author: samyakjain02@gmail.com}
\author{A. Bhagwat\\
ORCID: 0000-0002-3479-1301}
\email{Contributing author: ameeya@cbs.ac.in}
\affiliation{School of Physical Sciences, UM-DAE Centre for Excellence in Basic Sciences, University of Mumbai, Kalina Campus, Mumbai 400 098, India}

\date{July 8, 2024}

\begin{abstract}
 Tunneling half lives are obtained in a minimalistic deformation picture of nuclear decays \cite{samyak}. As widely documented in other deformation models, one finds that the effective mass of the nucleus changes with the deformation parameter. However, contrary to the approach used in literature, a position-dependant mass potentially makes using WKB tunneling probabilities unreliable for estimating nuclear lifetimes. We instead use a new approach, a combination of the Transmission Matrix and WKB methods, to estimate tunneling probabilities. Because of the simplistic nature of the model, the calculated lifetimes are not accurate, however, the relative trends in the lifetimes of isotopes of individual nuclei are found to be consistent. Using this, we develop an empirical scaling to obtain the actual half-lives, and find the primary scaling parameter to have remarkably consistent values for all nuclei considered. The new tunneling method proposed here, which produces very different probabilities as compared to the usual WKB approach, is another key result of this work, and can be utilized for arbitrary potentials and mass variations.
\end{abstract}

\maketitle
\newpage

\section{Introduction}
In a companion paper \cite{samyak} to this work, we developed a minimalistic macroscopic-microscopic nuclear deformation model, with the goal of preserving enough complexity to reproduce fission barriers with reasonable agreement, while being able to produce an analytical geometric picture of nuclear stability. To this end, we considered only the primary second order deformation parameter $a_2$, and truncated the obtained deformation potential to the cubic term. We were able to map this potential to the Fold Catastrophe, one of 11 unique geometric structures proposed in Thom's Catastrophe Theory \cite{thom}, to which any arbitrary function of 5 or less parameters can be mapped. We found that nuclear stability was directly correlated to the existence/depth of the local minimum (and correspondingly the potential barrier, see Fig.\ref{V_U238}) one obtains in the cubic deformation potential.

In this work, we examine whether tunneling through the obained potential barrier can account for nuclear lifetimes. In particular, we find that the effective mass of the nucleus varies with the deformation parameter, a feature seen in all such deformation models. However, most works utilize the WKB formalism to obtain tunneling probabilities, which is fundamentally incorrect since the entire formalism is based on a uniform mass. We thus propose a new method to calculating tunneling probabilities; we utilize a combination of the WKB formalism and the Transmission Matrix approach \cite{semicond}. The T-matrix approach accounts for a varying effective mass, but diverges near the crossing point, which is where we use WKB to bridge the gap. 

We find that the calculated lifetimes are very inaccurate (expected from the simplistic nature of the model), but the relative lifetimes between isotopes are found to be consistent. We develop an empirical scaling relation that relates the calculated and actual lifetimes, and find the primary scaling parameter to be remarkably consistent for the nuclei considered.

In Section \ref{model}, we briefly outline the procedure used to construct deformation model as in \cite{samyak}. In Section \ref{setting up} we set up the tunneling problem, and find the effective mass in the problem to be non-uniform. In Section \ref{new method}, we outline the Transmission matrix approach, and then propose a new tunneling method incorporating both WKB and the Transmission Matrix approaches. In Section \ref{results}, we present the comparison of the calculated and actual lifetimes of isotopes for four nuclei: Uranium, Plutonium, Protactinium, Neptunium (account for all even-odd combinations of $N$ and $Z$). We finally develop the aforementioned scaling relation, and find the primary scaling parameter to be consistent for the nuclei considered.

\section{Nuclear Deformation Picture}
\label{model}
The deformation model used here is motivated from the works of work of Nilsson (\cite{nilsson1}, \cite{nilsson2}, \cite{shapeandshells}), who semi-classically calculated fission barriers by parameterizing the energy of a nucleus in terms of its shape. This is done by first considering the nucleus as a spherical fluid drop; one can classically calculate the energy due to the volume due to inter-nucleon interactions, the loss of energy at the surface, and the Coulomb energy. One can then allow the nucleus to be deformed (symmetrically about an axis, say $z$, such that the volume is conserved) via a Legendre Polynomial expansion:
\begin{eqnarray}
    R(\theta) = R_\beta \left[ 1 + \left( \frac{2\lambda + 1}{4\pi} \right)^{1/2} \sum_{\lambda = 1} ^ \infty  \beta_\lambda P_\lambda (\theta)\right] \label{original legendre poly expansion}
\end{eqnarray}
where $R_\beta$ is obtained by demanding volume conservation. In the model considered, only the second deformation parameter, which has been shown to be the most important \cite{shapeandshells}, is considered:
\begin{eqnarray}
    a_2 = \sqrt{\frac{5}{4\pi}}\beta_2
\end{eqnarray}
Thus, the expansion simplifies to 
\begin{eqnarray}
    R_{a_2}(\theta) = R_0(a_2) \left[ 1 + a_2 P_2(\theta)\right] \label{legendre poly expansion}
\end{eqnarray}
One can then again calculate the extra surface and Coulomb energy the nucleus attains as a function of $a_2$; this is the classical deformation energy. 

This classical picture is found to explain nuclear binding energies well on average, but some nuclei, with particular values of $Z,N$, are found to be significantly more stable than expected. This naturally paints the picture of shells being filled by nucleons. One thus considers each nucleon to be in a harmonic potential caused by all other nucleons, and the corresponding eigen-energies are obtained. The nucleons are then filled in the orbitals as one does for electrons, but one cannot simply sum the nucleon energies to obtain the total nuclear energy. This is because this would lead to an over-counting of inter-nucleon interactions (since each nucleon is considered to be in a potential caused by all the other nucleons). One thus incorporates the nucleon eigenenergies as a perturbation to the classical energies; the mathematical details of this are described in [1]. \\

\begin{figure}[h]
\centering
\includegraphics[width=0.7\linewidth]{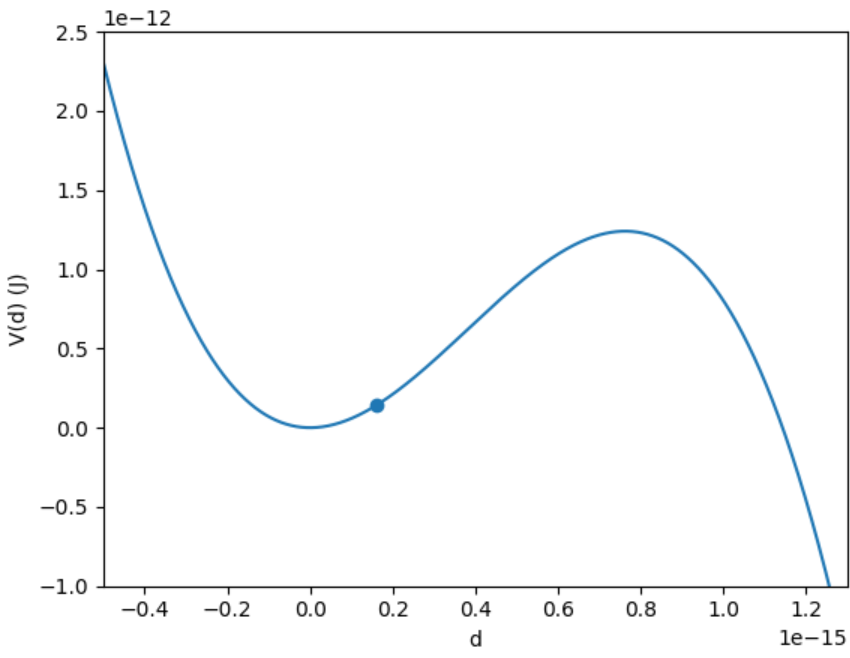}
 \caption{The deformation potential is plotted as a function of the deformation parameter $a_2$, for $U_{238}$. A potential barrier separating the local minimum from an unstable region is seen. The crossing point, with the potential equal to the zero-point energy (Eq.\ref{zero_pt}), is marked.} 

 \label{V_U238}
\end{figure}

In this semi-classical picture, one finally obtains a deformation potential of the form [1]:
\begin{eqnarray}
    V(a_2) = A a_2^3 + Ba_2^2 + Ca_2 \label{macro micro def energy}
\end{eqnarray}
where the expressions for $A(N,Z), B(N,Z), C(N,Z)$ are provided in [1]. For $U_{238}$, the deformation potential is plotted in Fig.\ref{V_U238}. 

 The potential immediately motivates a tunneling problem; can the tunneling probability through the barrier seen in Fig.\ref{V_U238} account for nuclear lifetimes?

\section{Setting up the tunneling problem with position-dependant mass}
\label{setting up}
 First, let us redefine the deformation parameter as
\begin{eqnarray}
    d = r_0 a_2
\end{eqnarray}
where $r_0 = 1.2 \times 10^{-15}$m is the Fermi radius, which is used to estimate the nuclear radius as
\begin{eqnarray}
    R(N,Z) = r_0A^{1/3}
\end{eqnarray}
This transformation gives the tunneling coordinate the dimensions of length. One can then re-scale $A,B,C$ appropriately to obtain the deformation potential as a function of $d$.

Akin to the Makri Miller description of semiclassical tunneling \cite{makri_miller}, we easily find the frequency $\omega$ near the minimum of the barrier as
 \begin{eqnarray}
     \omega = \frac{1}{M_0}\frac{\partial^2 V(d)}{\partial d^2}
 \end{eqnarray}
where $M_0$ is the effective mass of the nucleus, which we shall derive now.

For a nucleus with deformation parameter $a_2$, which is changing shape at a rate $\dot{a_2}$, we can classically estimate the nuclear kinetic energy to be
\begin{eqnarray}
    KE = \int_V (\rho dV) \frac{v(\vec{r})^2}{2}
\end{eqnarray}
where $\rho = \frac{Am_p}{\frac{4\pi}{3}R^3}$ is the nuclear density (assumed to be uniform). We can find the velocity at radial vector $\vec{r}$ as
\begin{eqnarray}
    v(\vec{r}) = \frac{r} {R_{a_2}(\theta)} \frac{d}{dt} R_{a_2}(\theta)
\end{eqnarray}
where $R_{a_2}(\theta)$ is the radial extent of the nucleus at polar angle $\theta$, and is given by Eq.\ref{legendre poly expansion}. By demanding volume conservation in Eq.\ref{legendre poly expansion}, we obtain
\begin{eqnarray}
    R_0(a_2) = \left( \frac{70}{4a_2^3 + 42a_2^2 + 70}\right)^{1/3} R
\end{eqnarray}
We can then substitute $R_0(a_2)$ in to the expressions for $R_{a_2}(\theta)$ and $v(\vec{r})$ to obtain, after some algebra, the kinetic energy as 
\begin{eqnarray}
    KE = \frac{M(a_2)}{2}(r_0\dot{a_2})^2
\end{eqnarray}
where
\begin{eqnarray}
    M(a_2) = \frac{3}{10 A^{5/3}} m_p h(a_2)
\end{eqnarray}
and 
\begin{eqnarray}
    h(a_2) = \frac{1}{R^5}\int d\theta \sin{\theta} R_{a_2}^2 \left( \partial_{a_2} R_{a_2} \right)^2
\end{eqnarray}
The final expression for $h(a_2)$ is obtained by symbolic evaluation in python.

Thus, the effective mass of the nucleus used to calculate the harmonic frequency must be the effective mass at the local minimum. The local minimum and maximum can be found by setting
\begin{eqnarray}
    \partial_{a_2}V(a_2) = 0
\end{eqnarray}
Let the minima and maxima obtained be
\begin{eqnarray}
    d_0 = r_0 a_2^{\text{min}}, d_m = r_0 a_2^{\text{max}}
\end{eqnarray}
Then, the harmonic frequency is given by 
 \begin{eqnarray}
     \omega = \frac{1}{M_0}\frac{\partial^2 V}{\partial d^2}(d_0)
 \end{eqnarray}

 The tunneling problem then simplifies to calculating the tunneling probability $P$ of a wavepacket with zero-point energy as 
 \begin{eqnarray}
     E = \frac{\hbar \omega}{2} + V(d_0) 
     \label{zero_pt}
 \end{eqnarray}
 One can then semi-classically obtain the half life of the nucleus as
 \begin{eqnarray}
     \tau = \frac{2\pi}{\omega} (\ln2) \frac{1}{P}
 \end{eqnarray}
where $P$ is the probability of a classical trajectory with the zero-point energy to tunnel through the barrier.

 \section{Tunneling with position-dependant mass}
 \label{new method}
The problem then reduces to calculating the transmission probability $P$. One can, of course, use the WKB approach by integrating the momentum through the barrier by simply incorporating the position dependant mass:
\begin{eqnarray}
    P = \exp \left({-2 \int d(r_0a_2) \left[ \frac{2M(a_2)}{\hbar^2}(V(a_2)-E) \right] ^{1/2}}\right)
\end{eqnarray}
However, this may not be valid; recall that the WKB probability is accurate only for small transmission coefficients. However, the above WKB expression of the mass is equivalent to discretizing the mass such that it is uniform individually in barriers of infinitesimal widths $d(r_0a_2)$, and summing up the integral: one is multiplying together the probabilities of transmission through each infinitesimal barrier. However, since the thickness of each barrier is small, the individual tunneling probabilities will be large, implying the WKB expression does not hold.
\subsection{Transmission matrix formalism}
An alternative to handle this non-uniform mass comes from transmission matrices \cite{semicond}, developed for condensed matter applications where the effective mass of electrons in separate junctions may be different. Let us consider a single potential step at $x = 0$, with the wavefunction on the left as 
\begin{eqnarray}
    \psi_1(x) = ae^{-ik_1x} + be^{ik_1x} \label{wavefn}
\end{eqnarray}
and on the right as 
\begin{eqnarray}
    \psi_2(x) = ce^{-ik_2x} + de^{ik_2 x}
\end{eqnarray}
One can impose the continuity of the wavefunction and its spatial derivative at $x = 0$, to obtain
\begin{eqnarray}
    a + b = c + d, \quad ik_1(a-b) = ik_2(c-d)
\end{eqnarray}
This can be easily rewritten as
\begin{eqnarray}
    \begin{pmatrix}
        c \\ d 
    \end{pmatrix} = T_{21}(0)\begin{pmatrix}
        a\\b
    \end{pmatrix}
    = \frac{1}{2k_2} \begin{pmatrix}
        k_2 + k_1 & k_2 - k_1 \\ k_2 - k_1&k_2+k_1
    \end{pmatrix}
    \begin{pmatrix}
        a \\ b
    \end{pmatrix}
\end{eqnarray}
where $T_{21}(0)$ is the transmission matrix between the regions $1$ and $2$ for a step at $x = 0$. For a step at $x = d$, one just needs to transform the spatial coordinate to $x-d$, after which $T(0)$ can be written. Then, one reverts back to the original coordinate to obtain \cite{semicond}
\begin{eqnarray}
    T_{21}(d) = \begin{pmatrix}
        e^{-ik_2d} & 0\\0&e^{ik_2d}
    \end{pmatrix}
    \frac{1}{2k_2} \begin{pmatrix}
        k_2 + k_1 & k_2 - k_1 \\ k_2 - k_1&k_2+k_1
    \end{pmatrix}
    \begin{pmatrix}
        e^{ik_1d} & 0\\0&e^{ik_1d}
    \end{pmatrix}
\end{eqnarray}

Now let us consider that both sides of the barrier have different effective masses $m_1$ and $m_2$. In such a case, the derivative matching condition is modified to \cite{semicond}
\begin{eqnarray}
    \frac{1}{m_1}\frac{d\psi(0^-)}{dx} = \frac{1}{m_2}\frac{d\psi(0+)}{dx}
\end{eqnarray}
one can verify that with this condition, the conservation of current is satisfied \cite{semicond}, and violated with the original condition.

One thus obtains
\begin{eqnarray}
    a+b = c+d,\quad i\frac{k_1}{m_1}(a-b) = i\frac{k_2}{m_2}(c-d)
\end{eqnarray}
With this, one obtains the transmission matrix at $x = d$ as
\begin{eqnarray}
    T(d) = \begin{pmatrix}
        e^{-ik_2d} & 0\\0&e^{ik_2d}
    \end{pmatrix}
    \frac{m_2}{2k_2} \begin{pmatrix}
        \frac{k_2}{m_2} + \frac{k_1}{m_1} & \frac{k_2}{m_2} - \frac{k_1}{m_1} \\ \frac{k_2}{m_2} - \frac{k_1}{m_1} & \frac{k_2}{m_2}+\frac{k_1}{m_1}
    \end{pmatrix}
    \begin{pmatrix}
        e^{ik_1d} & 0\\0&e^{ik_1d}
    \end{pmatrix}
\end{eqnarray}
To account for cases where $E<V(x)$ on any side of the barrier, one may replace the corresponding $k$ with $i\kappa$
\begin{eqnarray}
    \kappa = \left(\frac{2m}{\hbar^2}(V-E)\right)^{1/2}
\end{eqnarray}
Finally, one may obtain the transmission coefficient $t$ by setting $a = 1, b = r, c = t, d = 0$, where $t$ and $r$ are the transmission and reflection \textit{coefficients} respectively. One then obtains $t$ as
\begin{eqnarray}
    t = \frac{\det{T}}{T_{22}}
\end{eqnarray}
and the transmission \textit{probability} $P$ as
\begin{eqnarray}
    P = |t|^2 = \left| \frac{\det T}{T_{22}} \right|^2
\end{eqnarray}

To handle arbitrary potentials and mass variations then, one may discretize the potential and mass in infinitesimal potential steps ($V_j = V(x_j,\quad j = 0,1,2...n)$, calculate the transmission matrix in each barrier, and multiply them all together:
\begin{eqnarray}
    T = T_{n,n-1}T_{n-1,n-2}...T_{10}
\end{eqnarray}
and then calculate the transmission probability as above.

This approach seems quite powerful, in the sense that it does not make any approximations, unlike WKB, and should be able to handle arbitrary potentials and mass variations. However, this method fails near the crossing point, $E = V$. To see this, consider the wavefunction at a step on the crossing point: since the energy equals the potential in the step, the wavefunction is just a constant. As such, one cannot write the wavefunction as one did in Eq.\ref{wavefn}; one cannot define an incoming and reflected wave. Numerically, we found this translates to a failure of convergence of $T$ for an energy close enough to $V_j$ for any $j$. This has also been documented in \cite{diverg} \\

\subsection{A new tunneling method}
Thus, both WKB and the transmission matrix approach have flaws for such potentials; the WKB formalism is not valid for a varying effective mass, and the transmission matrix approach numerically diverges near the crossing points. We thus utilize a mixture of these approaches.

The transmission matrix approach is expected to be accurate some distance away from the crossing point. One can thus utilize the transmission matrix approach in the middle of the barrier (region $TR$) to compute some transmission probability $P_T$. On either side of this middle barrier, neither of these approaches work. However, since the mass variation in the regions are comparatively smaller, one would expect the error in the WKB approximation due to the varying mass to be correspondingly smaller. We thus utilize the WKB approach on both sides (regions $W_1, W_2$) of $TR$  up to the crossing points. One then has the transmission probability as 
\begin{eqnarray}
    P = P_{W_1} P_T P_{W_2}
\end{eqnarray}
To define the regions $TR, W_1, W_2$, we choose a threshold $t_h$, and cut off $W_1$ and $W_2$ at $x_1, x_2$ respectively, where $x_1, x_2$ are such that
\begin{eqnarray}
    \frac{V(x_1) - E}{E} = \frac{V(x_2) - E}{E} = t_h
\end{eqnarray}

\begin{figure}[h]
\centering
\includegraphics[width=0.9\linewidth]{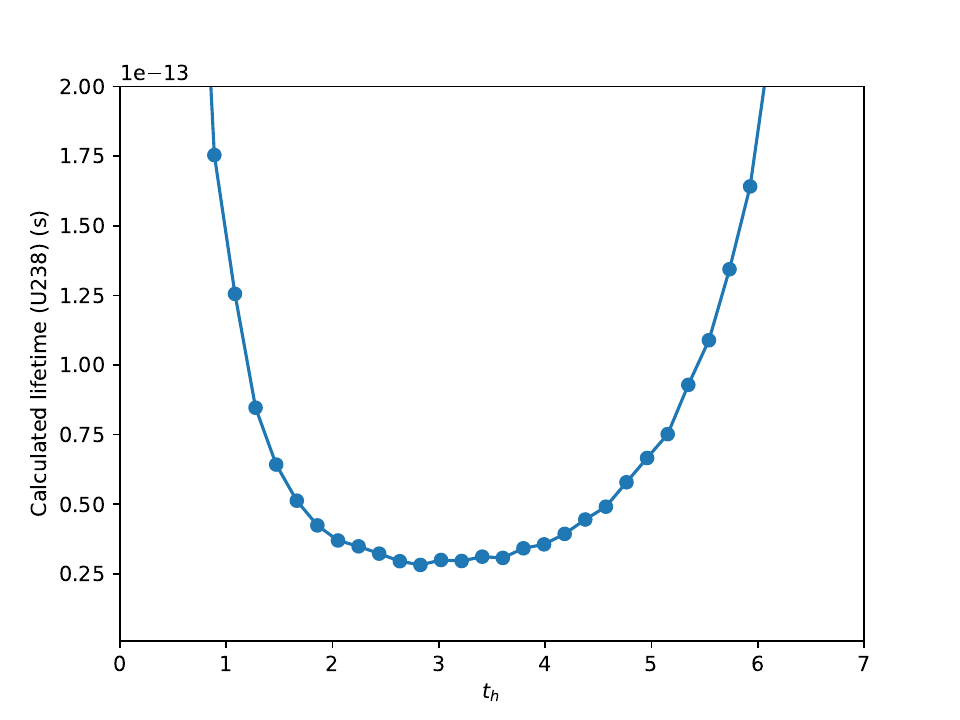}
 \caption{The calculated lifetime of $U_{238}$ is plotted as a function of the threshold $t_h$ via the proposed tunneling method. The threshold is chosen to be such that the lifetime does not change much with changes in the threshold, i.e, the local minimum of the plot.} 
 \label{threshold}
\end{figure}

Then all that remains is to choose a suitable threshold $t_h$. This can be done by demanding that the final tunneling probability should not be affected much by the choice of $t_h$. In Fig.\ref{threshold}, we plot the computed lifetime for $U_{238}$ as a function of the threshold; the threshold is chosen to be at the minimum of the plot. This is empirically observed to be approximately \begin{eqnarray}
    t_h \approx \frac{t_h^0}{2}
\end{eqnarray}
where $t_h^0$ is given by 
\begin{eqnarray}
    t_h^0 = \frac{V(d_m) - E}{E}
\end{eqnarray}
and, as defined earlier, $d_m$ is the local maximum of the potential considered.

\section{Results}
\label{results}
We can now check if our deformation model can produce any useful results in regards to the nuclear half-lives. We note here that the considered model was found to be valid in explaining $\alpha$ decays, but not $\beta$ decays \cite{samyak}. Thus, we shall only consider nuclei with significant $\alpha$ decay mode. Due to the simplicity of the model, we cannot hope to estimate the half-lives themselves accurately; it has been well documented that other deformation parameters need to be considered for this, which give rise to multiple barriers. Recall that the WKB probability goes as the exponential of the momentum integral; if the integral is off by a factor of half due to a missed barrier, the actual probability would be the square of the probability we obtain. This in turn leads to massive differences in the order of magnitudes of the half-lives we obtain. 

\begin{figure}[ht]
\begin{subfigure}{0.49\textwidth}
{\includegraphics[width = 1.05\linewidth]{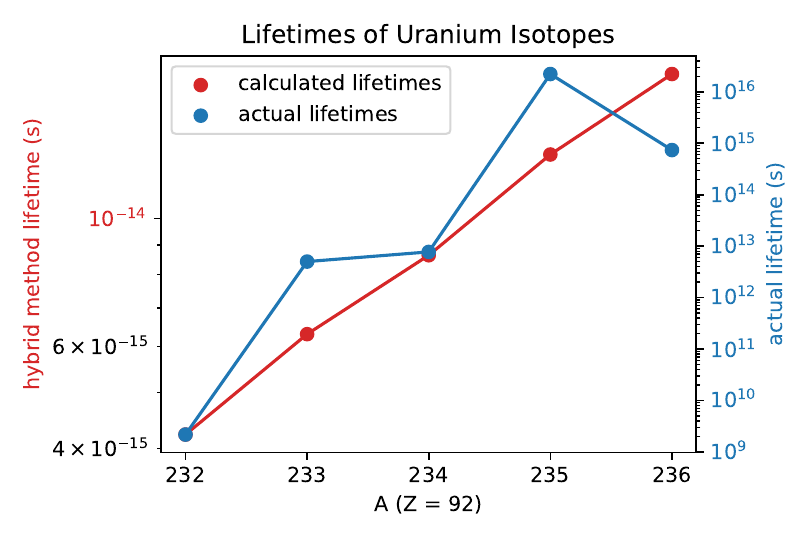}}
\subcaption{Uranium}
\end{subfigure} 
\begin{subfigure}{0.49\textwidth}
{\includegraphics[width = 1.05\linewidth]{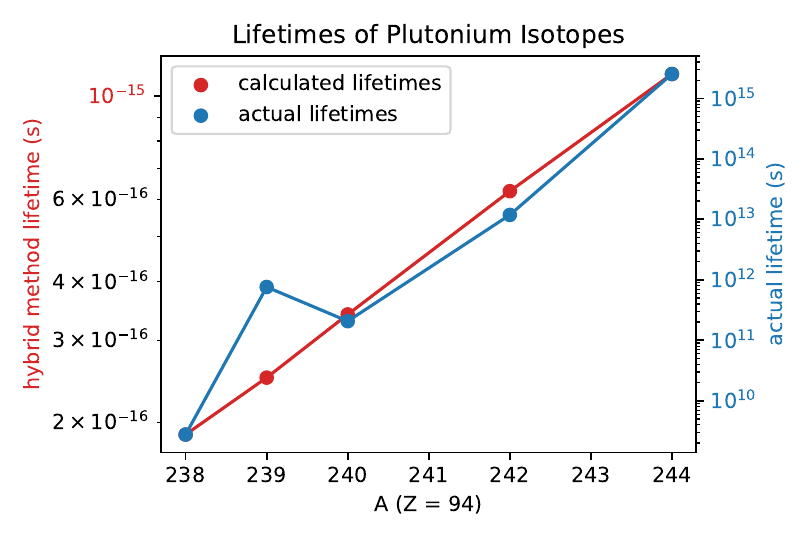}}
\subcaption{Plutonium}
\end{subfigure} 

\begin{subfigure}{0.49\textwidth}
\centerline{\includegraphics[width = 1.05\linewidth]{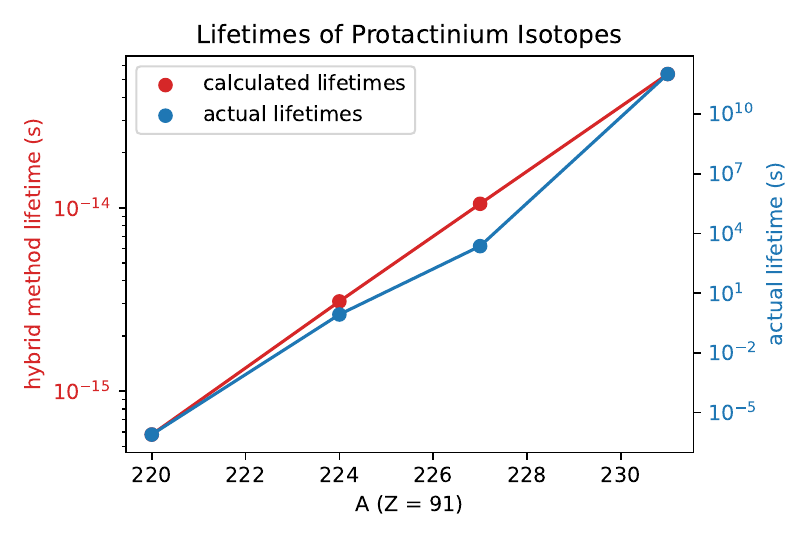}}
\subcaption{Protactinium}
\end{subfigure} 
\begin{subfigure}{0.49\textwidth}
\centerline{\includegraphics[width = 1.05\linewidth]{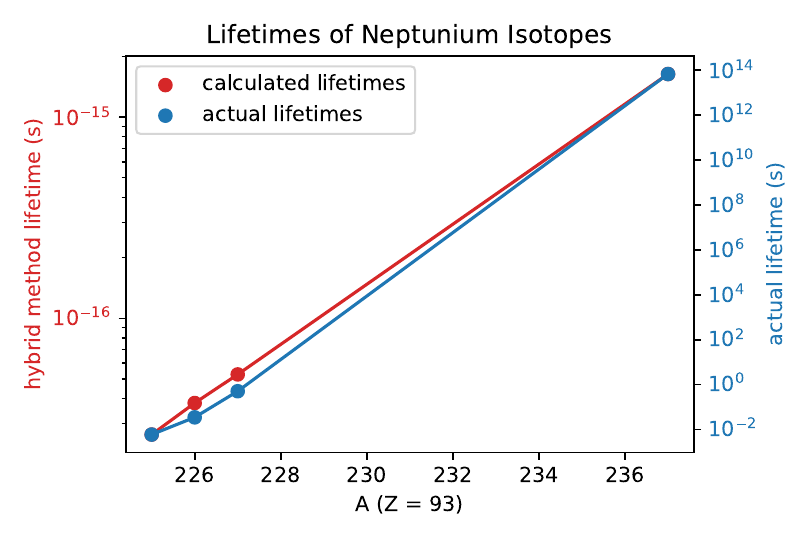}}
\subcaption{Neptunium}
\end{subfigure} 
\caption{Calculated and actual lifetimes of isotopes of various nuclei.}
\label{calc lives}
\end{figure}

Thus, we instead wish to see if we can estimate trends in nuclear half lives (for example, between isotopes), up to some scaling. In Figs.\ref{calc lives}, we plot the computed and actual lifetimes of various isotopes of Uranium, Plutonium, Protactinium, and Neptunium.

It is observed that the overall trends are captured quite well, with the appropriate scaling of the $y$ axes. Furthermore, the rough alignment of the plots (with different scaling of the $y$ axes) points towards an empirical relation of the form
\begin{eqnarray}
    \log{t_a} = m\log{t_c} + p(N,Z) \label{scaling}
\end{eqnarray}
where $t_a$ and $t_c$ are the actual and calculated lifetimes respectively. Here $p(N,Z)$ is included to account for the zig-zag nature of the actual lifetimes seen in the above figures, which can be attributed factor for odd or even number of neutrons in different isotopes. Thus we consider $p(N,Z)$ of the form

\begin{numcases}{p(N,Z)=}
  p_0, & \( N = \text{odd}, Z = \text{odd}\) \nonumber \\
  p_1, & \( N = \text{odd}, Z = \text{even}\) \nonumber \\
  p_2, & \( N = \text{even}, Z = \text{odd}\) \nonumber \\
  p_3, & \( N = \text{even}, Z = \text{even}\) \nonumber 
\end{numcases}

\begin{figure}[ht]
\begin{subfigure}{0.49\textwidth}
{\includegraphics[width = 0.99\linewidth]{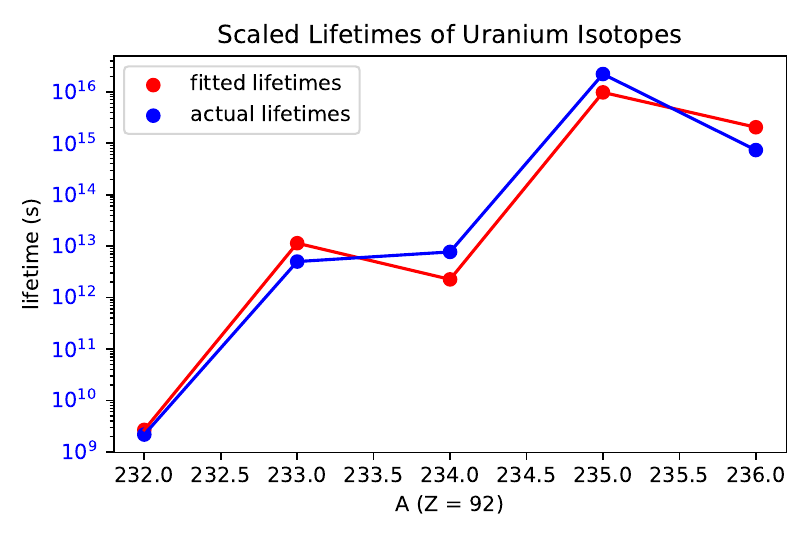}}
\subcaption{Uranium ($m = 9.413, p_2 = 146.7, p_2 = 144.7$)}
\end{subfigure} 
\begin{subfigure}{0.49\textwidth}
{\includegraphics[width = 0.99\linewidth]{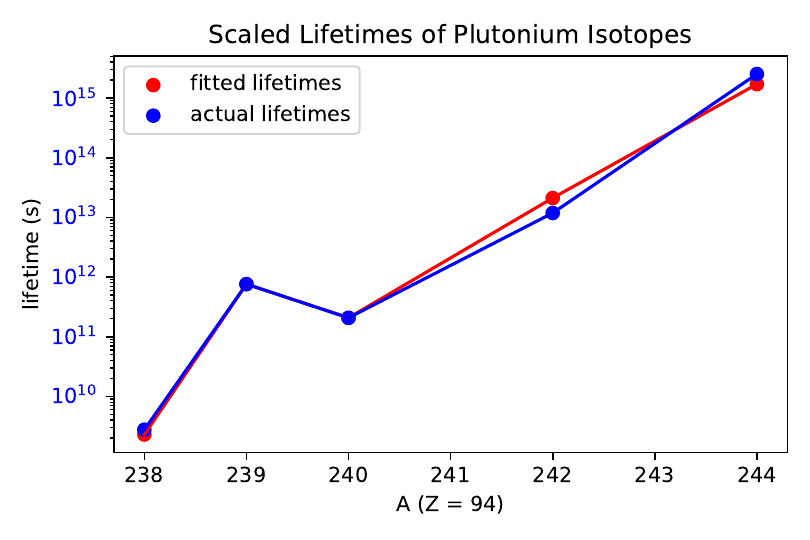}}
\subcaption{Plutonium ($m = 7.615, p_2 = 130.7, p_2 = 129.1$)}
\end{subfigure} 

\begin{subfigure}{0.49\textwidth}
\centerline{\includegraphics[width = 0.99\linewidth]{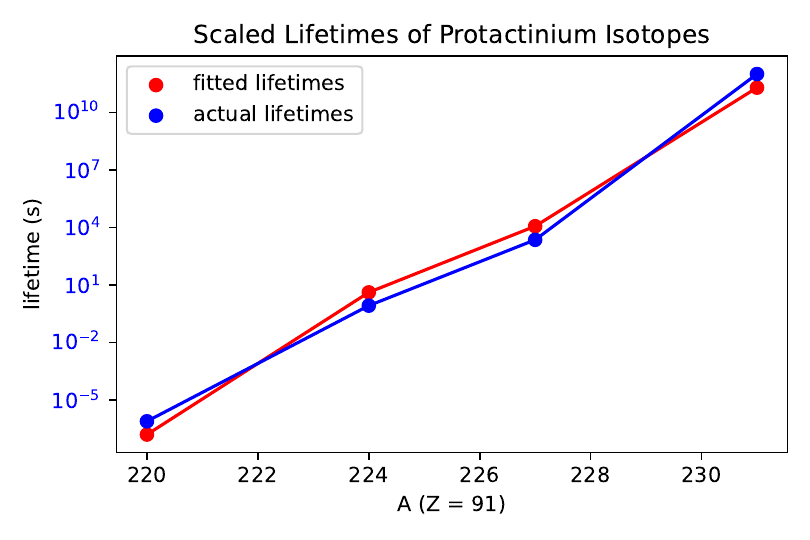}}
\subcaption{Protactinium ($m = 10.22, p_0 = 148.8, p_1 = 146.9$)}
\end{subfigure} 
\begin{subfigure}{0.49\textwidth}
\centerline{\includegraphics[width = 0.99\linewidth]{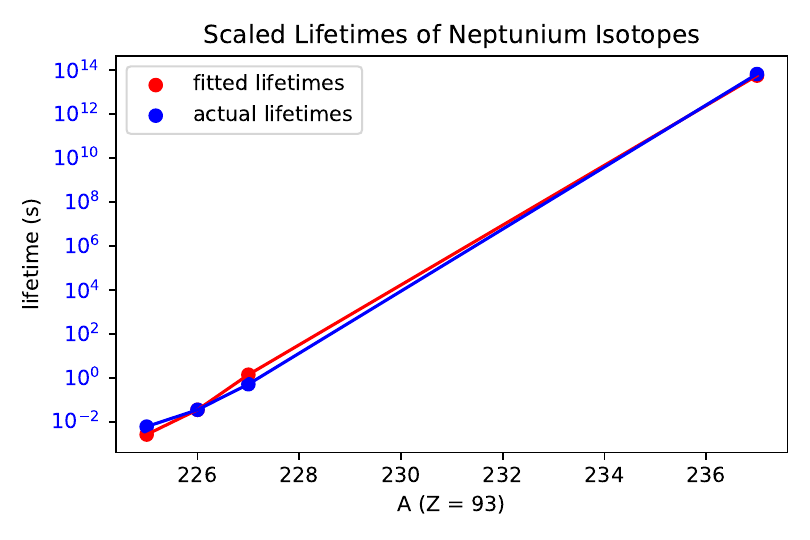}}
\subcaption{Neptunium ($m = 9.118, p_0 = 148.3, p_1 = 148.6$)}
\end{subfigure} 
\caption{Fitted and actual lifetimes along with their respective best-fit parameters.}
\label{final fit}
\end{figure}

For each $Z$, a best fit of $p_0,p_1,p_2,p_3$, and $m$, is performed. We find that the parameter $m$, which decides the relative lifetimes between isotopes, to be very consistent for all nuclei (average $m\approx 9.11$), with remarkable agreement with the actual half-lives in each fit individually. The fitted half lives and their corresponding the best-fit parameters are shown along with the actual half lives in Fig.\ref{final fit}.

\section{Conclusion}
Via a new proposed tunneling method, the considered model [1] is found to give accurate lifetimes (for nuclei with significant $\alpha$ decay mode participation) up to a logarithmic scaling defined by Eq.\ref{scaling}. The primary scaling parameter $m$ is found to be very similar for all nuclei considered, and approximately given by (average of values obtained here) $m \approx 9.11$. The other parameters show significant variation, and their average values are given by $p_0 \approx148.6, p_1 \approx 147, p_2 \approx 139.25, p_3 \approx 137.45$. 

The new proposed tunneling method utilizes a combination of WKB and Transmission Matrix theory. To verify its superiority over both in handling tunneling in arbitrary potentials (with varying mass), one must consider its application to similar problems with known analytical answers.

\section{Acknowledgements}
\vspace{-0.2cm}
We would like to thank Vikram Rentala and Kumar Rao for their valuable insights. Samyak Jain would like to thak Amber Jain, Paritosh Hegde and Reet Mhaske for their suggestions.

\textbf{Funding:} No funding was received for conducting this study.
\newline
\textbf{Conflict of interest:} The authors of this work declare that they have no conflicts of interest.


\bibliographystyle{unsrt}
\bibliography{main}
\end{document}